\documentstyle[aps,epsf,twocolumn]{revtex}
\begin{document}
\title{Comment on {\em Structure and Hyperfine Parameters of E\,$'_1$
Centers in $\alpha-$quartz and Vitreous SiO$_2$}}
\author{ Vincenzo Fiorentini and C. M. Carbonaro  }
\address{
Istituto Nazionale di Fisica della Materia  and
 Dipartimento di Scienze Fisiche, Universit\`a di Cagliari, 
via Ospedale 72, I-09124 Cagliari, Italy\\}
\date{\today}
\maketitle
\draft
Boero {\it et al.} have recently proposed \cite{loro} a model for the
E\,$'$ center in 
quartz and silica, based on ab initio studies of the
 singly positive charge state of the
oxygen vacancy V$_{\rm O}$. They find that 
in the +1 charge state, V$_{\rm O}$
 undergoes a distortion to a puckered configuration whose 
calculated hyperfine activity is consistent with the observed data. The 
purpose of this Comment  is to point out that
recent ab initio density-functional 
calculations on the oxygen vacancy in $\alpha-$quartz  \cite{noi}
 rule out  the possibility of identifying the +1 charge state  of the
 vacancy  with the  E\,$'$ center as suggested in \cite{loro}. 
Our results \cite{noi}  indicate
 the $Q=-3$ charge state  of V$_{\rm O}$  as another
possible candidate. 

\begin{figure}[ht]
\epsfclipon
\epsfxsize=6cm 
\epsfysize=4.5cm 
\centerline{\epsffile{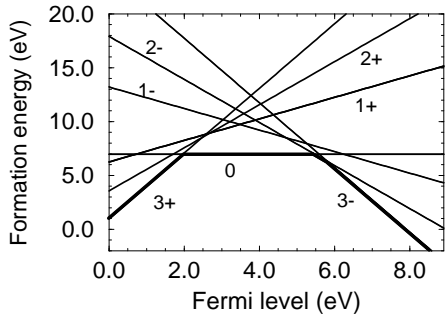}}
\caption{Calculated formation energies
 for the various charge states of the oxygen
vacancy in $\alpha-$quartz.}
\label{f-uno}
\end{figure}
The formation energies  of the oxygen  
vacancy   are displayed  in Fig.\ref{f-uno} as a function of the
 Fermi level (whose zero is chosen at the valence band top).
The oxygen vacancy is a   negative-$U$ center
whose   ground state can  have the  charge states
 $Q$=[+3,0,--2,--3]. 
 The singly charged states, in particular the +1
 state, are never stable against capture or release of further
electrons. Therefore the singly  positive vacancy
cannot be invoked as a candidate  E\,$'$ center.
This remains true accounting for the  puckering energy gain of 0.3 eV
\cite{loro}.

Our calculations are  technically quite  comparable to
those in \cite{loro}.
We studied a single oxygen vacancy in 36-atom $\alpha-$quartz
 supercells using conjugate-gradient total energy minimization
 with  ultrasoft pseudopotentials \cite{van}, a plane-waves  
basis cut off at   20 Ryd, and one special k-point.
 No symmetry restriction is imposed, and
all geometries are fully relaxed.  
 The theoretical lattice constants  (deviating 
less than 1\% from experiment)  are used. The energy zeroes in the
charged  supercells are  carefully aligned \cite{ac-gan-berlino}. 

\begin{figure}[ht]
\epsfclipon
\epsfxsize=6cm 
\epsfysize=4.5cm 
\centerline{\epsffile{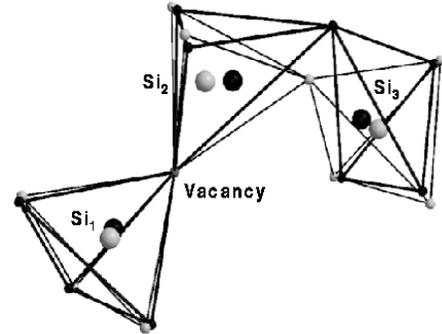}}
\caption{
Equilibrium configuration of V$_{\rm O}$ for $Q=-3$ 
(thick lines, black atoms) and  $Q=0$ (thin lines, grey atoms).}
\label{f-due}
\end{figure}

The $Q=-3$ state  of V$_{\rm O}$ is
 characterized by a major lattice  distortion
involving three  vacancy-neighboring
tetrahedral units (Fig.\ref{f-due}).  
While the $Q=0$ and $Q=+3$  states fail to match most of the magnetic
or optical properties of  the E\,$'$ center, the  $Q=-3$ state 
 satisfies most  requirements  for identification with
E\,$'$. Calculated optical transition energies agree well  with
absorption and emission bands  commonly associated with E\,$'$. The
main feature of the charge density of the  unpaired electron in the
defect state (not shown due to space constraints) is a pronounced lobe
associated with atom Si$_3$ (see Fig.\ref{f-due}), and pointing
approximately towards Si$_1$. The angle formed by this dangling lobe 
with the neighboring O atoms is $\sim$100$^{\circ}$ in accordance
with experimental estimates. Two additional, weaker  density 
features are associated with Si$_1$ and Si$_2$. The overall picture
seems compatible  with the observation of one strong,  and two weak
 hyperfine signals. Calculated hyperfine parameters are
unfortunately not available at the moment.

In summary  the +1 state of the oxygen vacancy in quartz cannot
 be associated with the E\,$'$ center.
An alternative candidate is  the
--3 charge state of the same center.

\end{document}